%% file: main.tex
\documentclass[aps,superscriptaddress,prl,nofootinbib,reprint]{revtex4-1}

\usepackage[ colorlinks = true,
             linkcolor = blue,
             urlcolor  = blue,
             citecolor = blue,
             anchorcolor = green,
]{hyperref}

\usepackage{tabularx}
\usepackage{amsfonts}
\usepackage{amssymb}
\usepackage{amsthm}
\usepackage{amsmath}
\usepackage{amsxtra}
\usepackage{graphics}
\usepackage{enumerate}
\input{variouscommands.tex}

\input{cftdefs.tex}

\usepackage{mathtools}
\mathtoolsset{showonlyrefs}

\begin{document}

\title{Matrix product approximations to multipoint functions\\
in two-dimensional conformal field theory}

\author{Robert K\"onig}
\affiliation{Institute for Advanced Study \& Zentrum Mathematik, Technische Universit\"at M\"unchen, 85748 Garching, Germany}
\email{robert.koenig@tum.de}

\author{Volkher B.~Scholz}
\affiliation{Department of Physics, Ghent University, 9000 Gent, Belgium}
\affiliation{Institute for Theoretical Physics, ETH Zurich, 8093 Z\"urich, Switzerland}
\homepage{http://users.ugent.be/~vscholz}

\begin{abstract}
Matrix product states (MPS) illustrate the suitability of tensor networks  for the description of interacting many-body systems: ground states of gapped $1$-D systems are approximable by MPS as shown by Hastings~\cite{Hastings:2007vw}.  In contrast,  whether MPS and more general tensor networks can accurately reproduce correlations in critical quantum systems, respectively quantum field theories,
has not been established rigorously. Ample evidence exists:  entropic considerations provide restrictions on the form of suitable Ansatz states, and numerical studies show that certain tensor networks can indeed approximate the associated correlation functions. 
Here we provide a complete positive answer to this question in the case of MPS and $2D$ conformal field theory: we give quantitative estimates for the approximation error when approximating correlation functions by MPS. Our work is constructive and yields an explicit MPS, thus  providing both suitable initial values as well as a rigorous justification of variational methods. 
\end{abstract}

\maketitle 


\emph{Introduction.}---%
Recent years have shown that the entanglement structure of quantum many-body 
systems is key to understanding their peculiar properties. An important tool in this line of research are Ansatz states built from tensor networks, such as Matrix Product States (MPS)~\cite{fannesnachtergaelewerner1992,Wolf:2007wt}, Projected Entangled Pair States~\cite{schuch2011classifying} or the Multi-scale Entanglement Renormalization Ansatz (MERA)~\cite{PhysRevLett.99.220405}. 
Beyond highly successful applications to models in condensed matter physics, various authors~\cite{2010PhRvL.105y1601H,2013PhRvL.110j0402H,jennings2015continuum,2010PhRvL.105z0401O,2010PhRvL.104s0405V} have started to apply tensor network  methods to quantum field theories. These studies
provide convincing (though mostly only numerical) evidence that tensor networks indeed constitute a valuable tool in understanding interacting quantum fields. Furthermore, it appears that these methods are able to surpass the main obstacles pointed out by Feynman~\cite{feynman1987difficulties} to applying the variational principle to quantum field theories. However, analytic results such as exact error bounds were missing so far,  in sharp contrast to the case of non-relativistic systems. A key result in the latter context is the work of Hastings~\cite{Hastings:2007vw} on tensor network approximations to 1D~gapped systems, which  constitutes the culmination of earlier work by various authors~\cite{Verstraete:2006ir,Schuchetalentropymps2008} and has led to concrete algorithms~\cite{LandauUmeshVidick14}.

In examining if tensor networks can be used to model the entanglement structure of quantum field theories, conformal field theories~\cite{francesco2012conformal} (CFTs) in 2D constitute a natural testbed: they describe a wide range of physical systems, and -- unlike many other instances of quantum field theories -- they can be constructed rigorously. Similarly, among the class of tensor network states,  MPS stand out due to their structural simplicity. For 1+1D~CFTs, the 1-dimensional structure of MPS appears to be particularly suitable. Following this reasoning, the use of  MPS in this context has previously been investigated from various viewpoints: recent studies have aimed to construct Ansatz MPS for quantum Hall systems~\cite{nielsen2012laughlin,nielsen2014bosonic,2013PhRvB..87p1112E,zaletel2012exact}, based on the commonly used assumption that such systems are described by 2D~CFTs. MPS have also been employed for the simulation of critical 1D quantum systems~\cite{PhysRevB.78.024410,2012PhRvB..86g5117P,2009PhRvL.102y5701P}. The empirical success of these studies, combined with the lack of explicit error bounds, are the main motivation for this work.

Prior work on describing spin systems by MPS relies heavily on the particularly simple entanglement structure of the latter: the entanglement entropy of an MPS is bounded by its so-called bond dimension. Casting the relationship between entanglement structure and suitability for variational physics in a quantitative form, it was shown~\cite{Verstraete:2006ir,Schuchetalentropymps2008} that approximability of a given many-spin quantum state by MPS is completely determined by the scalings of the R\'enyi-entropies of the associated reduced density matrices (such entropic arguments are also the basis of~\cite{Hastings:2007vw}). This may suggest that a similar 
entropy-based approach may apply to the question of approximability of CFTs by MPS by using the results from~\cite{Holzhey1994443,CalabreseCardy2009}. However,  such an argument would only yield qualitative information and not give rise to exact error bounds. In addition, it would require carrying out a lattice regularization of the CFT. Nevertheless, our work is guided by similar considerations and can be regarded as an algebraic version of these arguments. 

\emph{Our contribution.}---%
We present an algebraic approach to quantify the accuracy of MPS approximations to 2D~CFTs. As the latter describe critical $1D$~quantum systems, our work is similar in spirit to that of Hastings~\cite{Hastings:2007vw} (for non-critical systems), with the crucial difference that it is constructive: in addition to showing existence, our work provides an algorithm for explicitly building an approximating MPS, together with suitable error bounds. We expect that this explicit construction may serve as a useful starting point for numerical variational algorithms, e.g., in the description of critical quantum spin chains. It may also guide the way to establishing similar error bounds for related tensor network states such as the MERA. 
This provides the first analytic arguments supporting the fact that tensor networks can be used to model the entanglement structure of quantum field theories. 

\emph{Chiral CFTs.}---%
To state our results, let us first consider the case of a chiral CFT, which specifies  {\em correlation functions} commonly denoted by
\begin{align}
  \label{eq:correlationfunctions}
  f_{\varphi_1,\ldots,\varphi_n}(z_1,\ldots,z_n) = \langle \vphi_1(z_1) \cdots \vphi_n(z_n)\rangle \in \C\ .
\end{align}
Here the field $\varphi_j$ is inserted at $z_j\in \mathbb{C}$; the functions are defined on a suitable subset of the complex plane parametrizing a surface.  A key feature of the correlation functions~$f_{\varphi_1,\ldots,\varphi_n}$ is their transformation behaviour under conformal transformations~$z\mapsto \vartheta(z)\in\mathbb{C}$ applied to the arguments $(z_1,\ldots,z_n)$. This becomes particularly simple if all~$\varphi_j$'s belong to the special family of {\em primary fields}; this is the main case of interest, as other correlation functions may be deduced from the Ward identities, see e.g.,~\cite{francesco2012conformal}\footnote{The conformal transformation property of correlation functions can be taken as the defining feature to axiomatize CFTs: all objects of a CFT can in fact be constructed from such a family of functions~$\{f_{\varphi_1,\ldots,\varphi_n}\}_{n\in\mathbb{N}_0,\varphi_j\in\cF}$~\cite{Gaberdiel:2000br}.}.

Now recall (see e.g.,~\cite{Wolf:2007wt}) that an MPS is a state  on $(\C^d)^{\otimes n}$ defined through the existence of a finite set of matrices $A^j_{k} \in \mathsf{Mat}(\C^D)$, $k = 1,\ldots,d$, $j=1,\ldots,n$ such that $\ket{\sigma_{MPS}} = \sum_{k_1,\ldots,k_n} \tr[A^1_{k_1} \cdots A^n_{k_n}] \ket{k_1,\ldots,k_n}$. The parameter~$D$ is called the bond dimension, and determines the size of the Ansatz. We show how to construct matrices~$A^j_{k}$ such that the coefficients~$\tr[A^1_{k_1} \cdots A^n_{k_n}]$ of the MPS approximate correlation functions $f_{\varphi_1,\ldots,\varphi_n}(z_1,\ldots,z_n)$ of primary fields~$\{\varphi_j\}_j$ at equispaced insertion points~$z_j=j\cdot d+d_0$ in the complex plane and establish error bounds on the absolute difference~$\epsilon$. This provides a trade-off between the  bond dimension and the approximation accuracy: we show that the bond dimension scales inversely polynomial in the approximation error~$\epsilon$. That is, by increasing the bond dimension $D$ (see Table~\ref{tab:scaling} for the exact dependence of $D$ on the parameters), the error in the point-wise approximation of the correlation functions can be made arbitrarily small. Remarkably, the influence of the particular CFT on the scaling of the bond dimension can be reduced to a single parameter~$\dimctwo$ (which coincides with the dimension of the Lie algebra~$\mathfrak{g}$ in the case of WZW models considered below). Because this parameter determines the difficulty of approximation, it represents a measure of the entanglement encoded in the correlation functions.
{\renewcommand{\arraystretch}{2}\setlength{\tabcolsep}{1em}
\begin{table}[b]
  \begin{center}
    \begin{tabular}{ c | c | c }
      & fixed $n, d$ & fixed $\eps$\\
      \hline
      Scaling of $D$ & $\sim \left(\frac{1}{\eps}\right)^{\kappa \dimctwo \frac{n}{d}}$ & $\sim \gamma(\eps) e^{2 \pi\sqrt{\frac{1}{6}\dimctwo n}}$
    \end{tabular}
  \end{center}
  \caption{This summarizes the scaling of the bond dimension $D$ sufficient to obtain an approximation error $\eps$, as a function of the UV cutoff $d$ (i.e., the spacing of insertion points) and the number of insertions $n$. 
Here $\kappa$ and $\gamma(\eps)$ denote constants, and the parameter $\dimctwo$ encodes the dependence on the CFT. }
  \label{tab:scaling}
\end{table}}

We now review  some of the concepts necessary to establish our results. We subsequently outline a proof of our approximation result and conclude. 
 
\emph{Vertex operator algebras.}---%
The mathematical framework needed for our purposes is that of \emph{vertex operator algebras} (VOAs),  introduced by Borcherds~\cite{Borcherds:1992bi} as well as Frenkel~\cite{frenkel1989vertex} in their study of the representations of the Monster group. A VOA~$\cV$ is a complex vector space, together with a $z$-dependent multiplication rule, encoding the symmetries of the CFT. Let us illustrate this framework using the example of Wess-Zumino-Witten (WZW) models~\cite{wess1971consequences,witten1984non,novikov1981multivalued}, which we  employ for illustrative purposes throughout the remainder of the manuscript.  In addition to the conformal symmetries, these possess additional local symmetries given by a compact Lie group. This specific case was nicely worked out in~\cite{tsuchiya1987vertex,Frenkel:1992jt}, and we mainly rephrase their main arguments in physics terms. Starting point is a compact simple Lie algebra~$\g$, which is turned into an affine Lie algebra by employing the affinization $\ga = \g \otimes \C[t,t^{-1}]$ with the commutator rule $[\fa(n),\fb(m)] = [\fa,\fb](n+m) + \delta_{n,-m} n\, k\, \tr[\fa\fb]$ where $\fa(n) = \fa \otimes t^n$.  Here $\fa,\fb \in \g$, and $k$ is a real positive integer defining the \emph{level} of $\ga$. The VOA as a vector space is given by the full Fock space $\cV$ generated by the negative modes of $\ga$ (i.e., elements with negative exponent in $t$) acting on the vacuum vector $\1$ which is given by the identity element in $\g$.
Elements $\fa(n)$ in $\ga$ are identified with creation operators (with adjoint operators being the corresponding annihilation operators), by setting $\fa(n)\, (\fb_1(-m_1) \ldots \fb_k(-m_k)\1) = \fa(n) \fb_1(-m_1) \ldots \fb_k(-m_k)\1$. Free fields are defined as $\fa(z) = \sum_{n \in \Z} \fa(n) z^{-n-1}$.  In the VOA framework, the generalization of these fields to the entire Fock space is called the \emph{vertex operator} and its action on the Fock space equips the space with a $z$-dependent multiplication rule. 

The construction of general VOAs is more involved, however, our arguments remain valid in more general cases and we thus focus our presentation here on WZW models. The only assumption our results require is that we are dealing with a rational CFT which is generated by the modes of a finite-dimensional subspace. The latter condition was coined $C_2$-co-finiteness by Zhu~\cite{Zhu:1996cp}.

\emph{Modules and primary fields for WZW models.}---%
Given an irreducible representation of the Lie algebra $\g$ with highest weight $\lambda$, we can repeat the above Fock construction with the vacuum replaced by the associated highest weight vector $\vphi_\lambda$ resulting in a module $\modulekflambda{\lambda}$ for the WZW VOA. It possesses a natural $\Nl_0$-grading where the \emph{weight} $\wt$ of a vector $\fb_1(-m_1) \ldots \fb_k(-m_k)\vphi_\lambda$ is $\wt \lambda + \sum_{i} m_i$ with $\wt \lambda$ a positive number depending on $\lambda$. The module is irreducible if $\Scp{\theta}{\lambda} \leq k$, with $\theta$ the maximal root of $\g$. The set $\Lambda_k$ of such highest weights $\lambda$ is finite, implying that WZW theories are rational CFTs~\cite{francesco2012conformal}.

Conformal transformations are implemented by operators $L_m$, $m \in \Z$ satisfying the Virasoro commutation relations $[L_m,L_n]= (m-n) L_{n+m} + \frac{c}{12}\delta_{n,-m} (n^3-n)$ with $c$ being the central charge of the theory. These operators are obtained from creation and annihilation operators by means of the Sugawara-Segal construction~\cite{francesco2012conformal}. Of particular importance for us is the scaling operator (or free Hamiltonian) $L_0$, which is diagonal with respect to weight spaces (vectors of fixed weight) with eigenvalue being the weight.

Primary fields are defined as intertwiners between WZW modules: Let $V_{\lambda_i}$, $i=1,2,3$ be three irreducible highest weight representations of $\g$ with $\lambda_i \in \Lambda_k$. Turning the representation $V_{\lambda_1}$, $V_{\lambda_2}$ into irreducible WZW VOA modules $\modulekflambda{\vphi_i}$, $i=1,2$, a primary field is a linear mapping of elements $\vphi \in V_{\lambda_3}$ to linear $z$-dependent mappings $\vphi(z): \modulekflambda{\lambda_1} \to \modulekflambda{\lambda_2}$ such that we have commutation rule with creation operators $[\fa(n),\vphi(z)] = (\fa\vphi)(z)z^n$. 
We denote the set of these maps as $\cL(\modulekflambda{\lambda_{1}},\modulekflambda{\lambda_{2}})$. 
Correlation functions of WZW theories are constructed from a sequence of primary fields mapping $V_{\lambda_j}$ into $\cL(\modulekflambda{\mu_{j}},\modulekflambda{\mu_{j-1}})$, $j = 1,\ldots,n$, $\lambda_j, \mu_j \in \Lambda_k$ and elements $\vphi_j \in V_{\lambda_j}$, $v_0 \in \modulekflambda{\mu_{0}}$, $v_n \in \modulekflambda{\mu_{n}}$ as
\begin{align}
  \label{eq:correlationfunctionswzw}
  &f^{v_0,v_n}_{\vphi_1,\ldots,\vphi_n}(z_1,\ldots,z_n)=\Scp{v_0}{\vphi_1(z_1)\ldots \vphi_n(z_n) v_n}\,.
\end{align}
Expressing the fields in terms of their mode expression, $\vphi(z)=\sum_{m \in \Z} [\vphi]_m \,z^{-m-h_{\vphi}}$, the rhs.~of~\eqref{eq:correlationfunctionswzw} is seen to be a Laurent series in the indeterminates~$(z_1,\ldots,z_n)$. Here $h_\vphi = \wt \lambda_1 + \wt \lambda_3 - \wt \lambda_2$ is the \emph{scaling dimension}. Actual (complex) values of correlation functions are obtained on the domain of convergence by substituting complex numbers for these indeterminates. Vacuum correlation functions are obtained by choosing $\mu_0=\mu_n=0$, the adjoint module. While the fields based on creation and annihilation operators correspond to free fields, primary fields correspond to interaction between different particle realizations and hence are of particular interest.

\begin{figure}[t] 
  \includegraphics[scale=0.45]{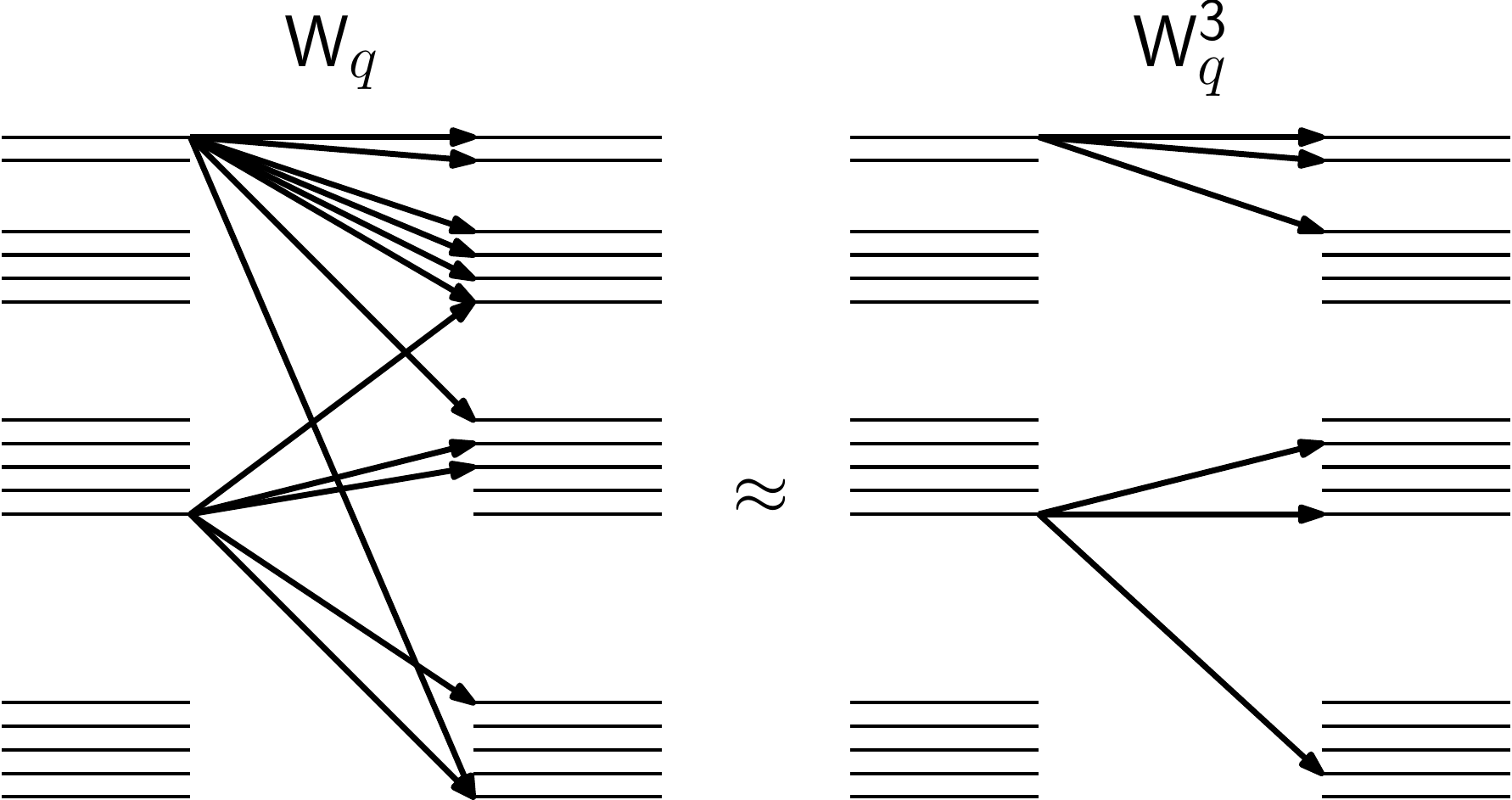}
  \caption{\emph{Illustration of the renormalization procedure:} (left) The regularized field operators $\W_q$ create arbitrary superpositions of energy eigenstates, hence an infinite amount of entanglement is needed to implement their action. The renormalized field operators $\W_q^N$ (right: $N=3$) only create superpositions of at most $N$ different energy eigenstates if applied to one element of the basis. Yet their action can be shown to be approximately the same as $\W_q$.}
  \label{fig:renorm}
\end{figure}

\emph{Proof idea.}---%
Our proof consists of two main steps: \emph{regularization} of the field operators and \emph{entanglement renormalization}. The correlation functions of the form~\eqref{eq:correlationfunctions} resemble the matrix element of a product of operators, and thus appear to be close to MPS form. However, the primary fields  have to be interpreted as Laurent series with coefficients in the set of linear operators between two modules, and constitute an unbounded operator. Our arguments start by noting that if the insertion points $z_1,\ldots,z_n$ are separated by a minimal distance, this deficiency can be taken care of and the field operators can be replaced by bounded operators. This is the \emph{regularization step}. However, the entanglement (respectively bond dimension) needed after regularization is still infinite. In order to proceed, we renormalise the fields by restricting their possibility to create superpositions of eigenmodes of the free Hamiltonian, see Fig.~\ref{fig:renorm}. This truncation procedure is closely related to the approximation by ``almost product states'' used in the exponential quantum de Finetti theorem~\cite{Konig_2009,Renner_2007}.

\emph{Proof for chiral WZW models.}---%
Let us now illustrate the key steps in our arguments in more detail, focusing for simplicity on vacuum correlation functions. In order to write a correlation function as in Eq.~\eqref{eq:correlationfunctionswzw} as a matrix element of \emph{bounded} operators, we first apply the following conformal mapping onto the insertions points: $\theta:z \to e^{-z}$. This transforms the equally spaced points on a line $z_j = d\cdot j$, $j=1,\ldots,n$, $d>0$ into $\theta(z_j) = q^j$ with $q=e^{-d} < 1$. By the assumption of conformal invariance, we get
\begin{align}
  &f^{\1,\1}_{\vphi_1,\ldots,\vphi_n}(\theta(z_1),\ldots,\theta(z_n)) =  q^{\sum_j \wt \lambda_j} f^{\1,\1}_{\vphi_1,\ldots,\vphi_n}(q,\ldots,q^n)\,.
\end{align}
Since the scaling transformation $z \mapsto qz$ is implemented by the operator $L_0$, $q^{L_0} \vphi(z) q^{-L0} = q^{\lambda}\vphi(q z)$ we get
\begin{align}
  f^{\1,\1}_{\vphi_1,\ldots,\vphi_n}(q,\ldots,q^n) = \Scp{\1}{\W_q^{(1)} \cdots \W^{(n)}_q \1}
\end{align}
where we introduced the regularized primary field $\W^{(j)}_q = q^{L_0}\vphi_j(1) q^{L_0}$. Following the results of Wasserman~\cite{Wassermann:1995fk}, regularized primary fields are bounded operators, $\norm{\W_q} \leq \bnd(q)$ with $\bnd(q)<\infty$ for $q<1$~\footnote{The exact form of the function $\bnd(q)$ depends on $\g$ and the modules involved in the definition of the primary field, but typically it scales as $\sim \frac{1}{1-q^s}$ for some $s\geq1$.}.
For general VOAs, the boundedness property follows from the fact that the trace of regularized primary fields exists, as shown by Zhu and Huang~\cite{Zhu:1996cp,Huang:2005gs}. This constitutes the regularization step.

The regularized field operator is bounded, but still acting on a infinite-dimensional Hilbert space. In order to obtain a finite bond dimension, we thus need an appropriate approximation. It is constructed by considering the mode expansion
\begin{align}
  \W_q = q^{L_0}\vphi(1) q^{L_0} = \sum_{m \in \Z} \,q^{L_0} [\vphi]_m q^{L_0} 
\end{align}
and discarding all modes $[\vphi]_m$ which change the grading of the module by more than $N \in \Nl_0$, the truncation parameter. We denote this renormalized primary field by $\W_q^N$. Acting with $\W_q^N$ on any eigenstate of the free Hamiltonian $L_0$ only creates a finite superposition of energy eigenstates (cf. Fig~\ref{fig:renorm}), hence only a finite amount of entanglement is needed to cope with the insertion of renormalized operators.
As a next step, we have to bound the error obtained from replacing the operator $\W_q$ by its truncated version $\W_q^N$. For that, let $v \in \modulekflambda{\mu}$ be an arbitrary vector of norm one which we decompose into orthogonal components $v = \sum_m v_m$ with $L_0 v_m= (m+\wt \mu) v_m$ and consider the norm difference $\norm{\W_q v-\W_q^N v}$. Since $\sqrt{q}^{-\wt \lambda} \W_q = \sqrt{q}^{L_0} \W_{\sqrt{q}} \sqrt{q}^{L_0}$ we have by the Cauchy-Schwarz inequality and the properties of the operator norm
\begin{align}
  &\norm{\W_q v-\W_q^N v}^2 \leq \\
  &\;\sum_m \left(\norm{(\idty -P_{[m-N,m+N]})\sqrt{q}^{L_0}} \cdot \sqrt{q}^m \right)^2 \cdot \bnd(\sqrt{q})^2\,,
\end{align}
where $P_{[m-N,m+N]}$ is the projection onto all states with weight in the interval $[m+\wt\mu-N,m+\wt \mu+N]$. We are left to bound the first term in the bracket. Starting from a vector $v_p$ of fixed weight $p \in \Nl_0 + \wt \mu$, we have $\norm{(\idty -P_{[m-N,m+N]})\sqrt{q}^{L_0}v_p} = \sqrt{q}^p \norm{v_p}$ if $p \notin [m-N,m+N]$. Applying this identify to an arbitrary vector $v=\sum_p v_p$ we find $\norm{(\idty -P_{[m-N,m+N]})\sqrt{q}^{L_0}v}^2 \leq (1-q)^{-1} q^{m+N+1}$ if $m\leq N$ and an additional summand of $(1-q)^{-1}$ otherwise. Summing these estimates over $m$ yields 
\begin{align}
  \label{eq:errorestimate}
  \norm{\W_q v-\W_q^N v} \leq q^{N/4}\frac{\sqrt{3} \bnd(\sqrt{q})}{1-\sqrt{q}}\,,
\end{align}
providing us with a norm bound for replacing the regularized field operator with its renormalized version. As expected, it is particularly useful for large $N$ and large $d$, since $q=e^{-d}<1$. By employing the telescoping sum technique, this estimate can be iterated to obtain an error estimate for the difference between the correlation function $\Scp{\1}{\W_q^{(1)}\cdots\W_q^{(n)}\1}$ and its renormalized version $\Scp{\1}{\W_q^{(1),N}\cdots\W_q^{(n),N}\1}$. Since this operator $\W_q^N$ only changes the weight by $N$, its action on any fixed vector, in particular the vacuum $\1$, gives rise to a finite-dimensional Hilbert space, and the same argument applies to the sequence $\W_q^{(1),N}\cdots\W_q^{(n),N}$, mapping $\1$ into the subspace $P_{[nN]}\modulekflambda{0}$ of vectors of weight $n\cdot N$ or less. Projecting each renormalized field onto $P_{[nN]}\modulekflambda{0}$ leaves the renormalized correlation function invariant and gives rise to a matrix on a finite dimensional Hilbert space. 

We obtained the approximation of correlation functions by matrix elements of operators on a finite-dimensional Hilbert space, and hence of MPS form. In order to estimate the dimension $D$ of this bond Hilbert space we have to bound the number of states of weight $nN$ or less. For WZW models, the number of states at each level $m$ (states with weight $m+\wt \mu$) can be upper bounded by the number $p(m,\dim \g)$ of $\dim \g$-component multi-partitions of $m$~\footnote{A partition $\mu$ of an integer $n$ is a set of integers $\mu = \{\mu^1,\ldots,\mu^l\}$ which sum up to $n$, $|\mu| = \sum_i \mu^i = n$. An $m$-component multi-partition of an integer $n$ is a set $\{\mu_1,\ldots, \mu_m\}$ of $m$ partitions such that $\sum_i |\mu_i| = n$.}. Adapting a proof strategy of Siegel (see~\cite{apostol2013introduction} for an exposition), 
we find $\log p(m,\dim \g) \leq 2 \pi \sqrt{\frac{\dim \g \cdot m}{6}}$ and hence 
\begin{align}
  \label{eq:bonddim}
  D \leq \sum_{m\leq nN} p(m,\dim \g) \leq \dim \g \cdot n\cdot N \cdot e^{2\pi \sqrt{\frac{n\cdot N\cdot\dim \g}{6}}}\,.
\end{align}
The bond dimension scales sub-exponentially with the truncation parameter $N$ while the error estimate obtained from iterating Eq.~\eqref{eq:errorestimate} decreases exponentially in $N$. Combining both facts gives the relationship between the growth of the bond dimension and the approximation error summarized in table~\ref{tab:scaling}.

\emph{Full CFTs.}---%
For full CFTs, correlation functions depend on additional ``conjugate'' parameters $(\bar{z}_1,\ldots,\bar{z}_n)$: this roughly corresponds  to gluing two copies of a chiral theory together. The Hilbert space is the direct sum $\cH = \oplus_{\lambda \in \Lambda_k} \modulekflambda{\lambda} \otimes \overline{\modulekflambda{\lambda}}$ where $\overline{\modulekflambda{\lambda}}$ denotes the complex conjugated Fock space, and primary fields correspond to linear combinations of tensor products of a chiral primary field and its anti-chiral (complex-conjugated) version. The regularization and renormalization steps outlined in the chiral case can be repeated analogously by doubling each expression to also involve the anti-chiral part. This yields a finite-dimensional operator of the form $\sum_i A_i \otimes \overline{A_i}$, which can be identified with a completely positive (CP) map on matrices, or transfer operator in MPS language. We conclude that vacuum correlation functions of full WZW models can be approximated by expectation values~$\tr[\Omega\,\mathbb{E}_1\circ\cdots\circ\mathbb{E}_n(\Omega)]$ of sequences of CP maps $\mathbb{E}_1,\ldots,\mathbb{E}_n$, where $\Omega$ is the matrix corresponding to the full vacuum. This is the defining form of finitely correlated states~\cite{fannesnachtergaelewerner1992,Wolf:2007wt} (FCS), a generalization of MPS to mixed states.


\emph{Conclusions and Outlook.}---%
Our  results  show that MPS tensor networks approximate general correlation functions of a CFT arbitrarily well.
The extension of our arguments to general VOAs and hence CFTs requires quite a few additional technical steps, but can be achieved as discussed in~\cite{fullversion}. Although we focused here on MPS tensor networks, we believe that our techniques are also applicable to study other tensor network Ansatz classes, such as MERA. Another question deserving further study is how our results fit in the renormalization framework developed by B\'{e}ny and Osborne~\cite{Beny_2015}. 


\emph{Acknowledgments.}---%
We thank Patrick Hayden, Tobias Osborne and Michael Walter for helpful discussions. RK  is supported by the Technische Universit\"at M\"unchen -- Institute  for  Advanced  Study,  funded  by  the  German  Excellence  Initiative  and  the European  Union  Seventh  Framework  Programme  under  grant  agreement  no.~291763. VBS is supported by the EC through ERC Qute. 


\bibliographystyle{apsrev4-1}
\bibliography{q}

\end{document}

%% file: variouscommands.tex
\def\idty{{\leavevmode\mathbf{I}}} 
\def\C{{\mathbb C}}     
\def\Nl{{\mathbb N}}     

\def\norm #1{\Vert #1\Vert}

\def\ket #1{\vert #1\rangle}

\def\tr{{\rm Tr}}
\newcommand\Scp[2]{\ensuremath{\, \langle #1 , #2 \rangle}}

\newcommand*{\vphi}{\varphi}

\newcommand*{\cF}{\mathcal{F}}

\newcommand*{\cH}{\mathcal{H}}

\newcommand*{\cL}{\mathcal{L}}

\newcommand*{\cV}{\mathcal{V}}

\newcommand{\bc}{\begin{center}}
\newcommand{\ec}{\end{center}}

\def\Z{\mathbb{Z}}

\def\01{\{0,1\}}

\newcommand{\eps}{\varepsilon}

\newcommand*{\modulekflambda}[1]{\mathsf{L}_{k,#1}} 


%% file: cftdefs.tex
\newcommand{\dimctwo}{C_{\cV}}

\newcommand*{\1}{\mathsf{1}}
\newcommand*{\wt}{\mathsf{wt}\,}

\newcommand{\W}{\mathsf{W}}

\newcommand*{\g}{\mathfrak{g}}
\newcommand*{\ga}{\hat{\mathfrak{g}}}

\newcommand*{\fa}{\mathfrak{a}}
\newcommand*{\fb}{\mathfrak{b}}

\newcommand*{\bnd}{{\vartheta}} 

